# THE LEP LEGACY

G. GIACOMELLI & R. GIACOMELLI

*Physics Dept., University of Bologna, INFN, Sezione di Bologna*
*Viale B. Pichat 6/2, BOLOGNA, I-40127 Italy*
*Email giacomelli@bo.infn.it, giacomellir@bo.infn.it*

In this lecture we shall summarize the scientific legacy of LEP, in particular in connection with the Standard Model of Particle Physics; we shall also discuss some historical and sociological aspects of the experimentation at LEP.

## 1. Introduction

In order to better understand the LEP contribution to particle physics it is appropriate to briefly recall the main features of the Standard Model (SM) of the Electroweak (EW) and Strong (SI) Interactions [1]. In this theory the fundamental constituents of matter are the quarks and the leptons, which may be considered as pointlike. The quarks and leptons appear in 3 families (generations), each made of 2 quarks and 2 leptons, one neutral and one negatively charged. The first family consists of the quarks u, d and the leptons $\nu_e$, $e^-$. The second family is composed of the quarks charm (c) and strange (s) and by the leptons $\nu_\mu$ and $\mu^-$; the third family is composed of the quark top (t) and bottom (b) and by the leptons $\tau^-$ and $\nu_\tau$. According to the Strong Interaction theory (QCD) each quark comes in three colours, green, red and blu. The only difference between families is the mass which becomes progressively larger when going from the first to the second and third family [and the lifetimes become smaller]. The SM does not explain why there are 3 families. In the SM to each quark and lepton corresponds an antiquark and an antilepton.

Quarks and leptons are subject to the EW Interaction mediated by the photon and the weak intermediate bosons $W^+$, $W^-$, $Z^0$. The SI between coloured quarks is mediated by 8 gluons (the leptons are not subject to the SI).

The formal structure of the SM is based on the gauge symmetry, which requires zero masses for quarks and leptons. In order to explain the observed masses, we introduce at least one scalar Higgs boson, which is needed for the spontaneous breaking of the symmetry and the generation of masses: the observed masses are due to the interaction of the Higgs boson with quarks and leptons. The Higgs also accounts for the large masses of the intermediate vector bosons. The coupling of the Higgs boson is predicted by the SM, but not its mass.

LEP yielded a very large number of important experimental results (see the Particle Data Books) and has placed the SM on a solid experimental ground.

The Large Electron Positron collider (LEP) was housed in a 27 km tunnel at ~100 m underground [2]. In 4 locations were placed the experiments ALEPH, DELPHI, L3 and OPAL. LEP took data from 1989 till 1995 at c.m. energies of ~91 GeV (LEP1) and from 1995 to 2000 at energies of 130-209 GeV (LEP2).

High energy colliders allow to study particle collisions at the highest energies, since the c.m. energy grows linearly with beam energy $E_b$, $E_{cm}=2E_b$. $e^+e^-$ collisions allow the best study of the fundamental particles and of their interactions.

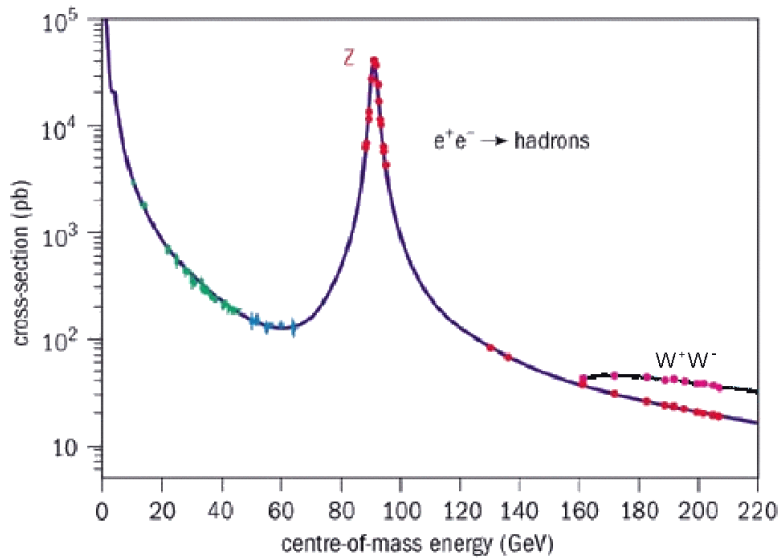

Fig.1. Hadron production cross section for $e^+e^- \rightarrow q\bar{q} \rightarrow$ hadrons vs c.m. energy.

Besides energy, another important parameter of a collider is its luminosity L, defined as that number which multiplied by a cross-section $\sigma$ gives the collision rate N: $N=L\sigma$. LEP had luminosities of $10^{31}$-$10^{32}$ $cm^{-2}s^{-1}$ which yielded collision rates of ~1 event/s at LEP1 and ~0.01 event/s at LEP2. Recent lower energy $e^+e^-$ factories have much larger luminosities.

Each of the 4 LEP experiments was a nearly $4\pi$ general purpose detector, made of many subdetectors. Their combined role was to measure the energy, direction, charge, and type of each produced particle. Apart from neutrinos and neutralinos, no particle was able to escape without leaving some sign of its passage. Each subdetector had a cylindrical structure with a "barrel" and two

"end-caps". Tracking was performed by a central detector; electron and photon energy measurements were carried out by electromagnetic calorimeters; the magnet iron yoke was instrumented as a hadron calorimeter and was followed by a muon detector [3]. A forward detector completed the e.m. coverage and was used as a luminosity monitor. The quality of the detectors and the relatively low event rate allowed to study in detail each event.

Fig.1 shows a compilation of data on $e^+e^- \to$ hadrons up to the highest LEP energies: up to 60 GeV the cross section decreases smoothly, then it is dominated by the $Z^0$; at higher energies it decreases, and above 160 GeV there is a structure connected with the opening up of the $e^+e^- \to W^+W^-$ channel.

## 2. Precision electroweak measurements

At energies around the $Z^0$ peak the basic processes are

$$e^+e^- \to Z^0, \gamma \to f\bar{f}, \quad f\bar{f} = q\bar{q}, l\bar{l}. \quad (1)$$

| | Measurement | Fit | $|O^{meas}-O^{fit}|/\sigma^{meas}$ 0  1  2  3 |
|---|---|---|---|
| $\Delta\alpha_{had}^{(5)}(m_Z)$ | 0.02761 ± 0.00036 | 0.02770 | |
| $m_Z$ [GeV] | 91.1875 ± 0.0021 | 91.1874 | |
| $\Gamma_Z$ [GeV] | 2.4952 ± 0.0023 | 2.4965 | |
| $\sigma_{had}^0$ [nb] | 41.540 ± 0.037 | 41.481 | |
| $R_l$ | 20.767 ± 0.025 | 20.739 | |
| $A_{fb}^{0,l}$ | 0.01714 ± 0.00095 | 0.01642 | |
| $A_l(P_\tau)$ | 0.1465 ± 0.0032 | 0.1480 | |
| $R_b$ | 0.21630 ± 0.00066 | 0.21562 | |
| $R_c$ | 0.1723 ± 0.0031 | 0.1723 | |
| $A_{fb}^{0,b}$ | 0.0992 ± 0.0016 | 0.1037 | |
| $A_{fb}^{0,c}$ | 0.0707 ± 0.0035 | 0.0742 | |
| $A_b$ | 0.923 ± 0.020 | 0.935 | |
| $A_c$ | 0.670 ± 0.027 | 0.668 | |
| $A_l$(SLD) | 0.1513 ± 0.0021 | 0.1480 | |
| $\sin^2\theta_{eff}^{lept}(Q_{fb})$ | 0.2324 ± 0.0012 | 0.2314 | |
| $m_W$ [GeV] | 80.425 ± 0.034 | 80.390 | |
| $\Gamma_W$ [GeV] | 2.133 ± 0.069 | 2.093 | |
| $m_t$ [GeV] | 178.0 ± 4.3 | 178.4 | |

Table 1. Precision measurements of the electroweak parameters obtained by the global fit of all the data from the four LEP experiments [4].

The $q\bar{q}$ pairs are $u\bar{u}$, $d\bar{d}$, $s\bar{s}$, $c\bar{c}$, $b\bar{b}$. Each q or $\bar{q}$ hadronizes (fragments) into a jet of hadrons. The $l\bar{l}$ pairs are charged ($e^+e^-$, $\mu^+\mu^-$, $\tau^+\tau^-$) or neutral ($\nu_e\bar{\nu}_e$, $\nu_\mu\bar{\nu}_\mu$, $\nu_\tau\bar{\nu}_\tau$).

The behaviour of the cross-section around the $Z^0$ peak is typical of a resonant state with J=1, and is well described by a relativistic Breit-Wigner formula plus electromagnetic and interference terms. The formula has to be convoluted with initial state radiation. Around the $Z^0$, the last 2 terms are small corrections to the main $Z^0$ Breit-Wigner term. The formula depends on $m_Z$ and on the partial width for the $Z^0$ decay into a fermion-antifermion pair. The total width $\Gamma_Z$ is given by

$$\Gamma_Z = \Gamma_h + \Gamma_e + \Gamma_\mu + \Gamma_\tau + N_\nu \Gamma_\nu = \Gamma_{vis} + \Gamma_{inv} \qquad (2)$$

where $\Gamma_h$ is the hadronic width and $\Gamma_e$, $\Gamma_\mu$, $\Gamma_\tau$, $\Gamma_\nu$ are the leptonic widths.

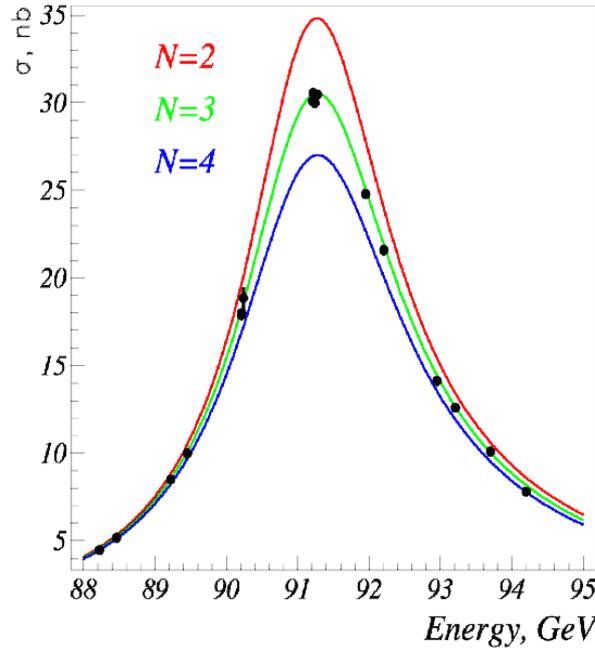

Fig.2. The shape of the $Z^0$ resonance yields information on the number of light neutrino types [three and only three].

At each energy around the $Z^0$ peak, measurements of the cross-sections were made for $Z^0 \to$ hadrons, $\to e^+e^-$, $\to \mu^+\mu^-$, $\to \tau^+\tau^-$, the forward-backward lepton asymmetries $A_{FB}^e$, $A_{FB}^\mu$, $A_{FB}^\tau$, the $\tau$ polarization $P_\tau$, the $b\bar{b}$ and $c\bar{c}$ partial widths and forward-backward asymmetries, and the $q\bar{q}$ charge asymmetry.

To combine results from the 4 LEP experiments, each experiment provided a set of optimized parameters (at beginning 9 parameters, 5 if lepton universality is assumed). Later more parameters were added. The latest ones are $m_W$, $\Gamma_W$.

Many parameters are expressed in terms of the effective EW mixing angle

$$\sin^2 \theta_{\text{eff}}^l = \frac{1}{4}(1 - \frac{g_v}{g_a}). \qquad (3)$$

The present best values of this and other parameters are given in Table 1.

The properties of the $Z^0$ have been studied with great precisions by the 4 experiments [4]: the $Z^0$ mass is now known with a precision of 2 parts in 100000 and the lifetime to 0.1 %. The measurements of the couplings of the $Z^0$ to quarks and leptons are tests of the SM to 0.1 %. Results were also obtained assuming lepton universality, which seems to be well established at LEP. The interactions of the $Z^0$ are those predicted by the gauge symmetry, while the masses do not reflect the symmetry. This is an indication for the Higgs mechanism.

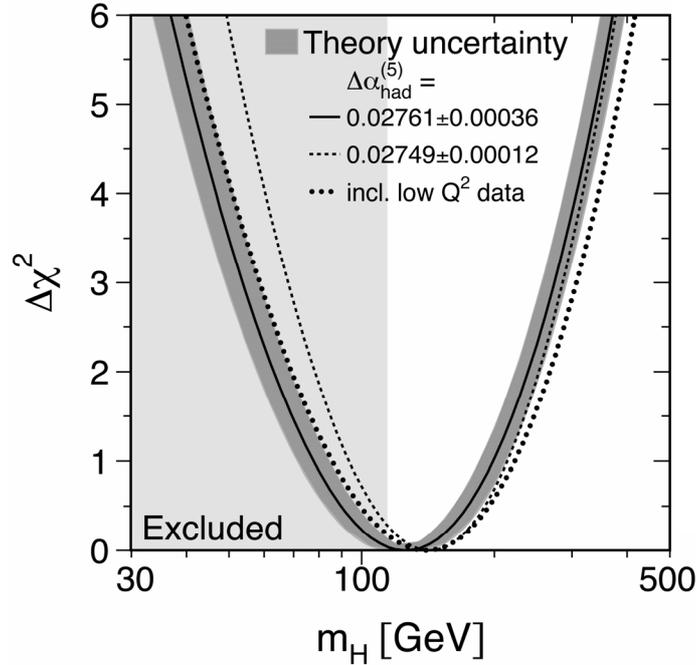

Fig.3. Fit probability vs $m_H$. The solid line is the result of the fit using all EW data; the band is an estimate of the theoretical error due to missing higher order corrections. The vertical line at $m_H$=115 GeV is the 95% C.L. limit from direct searches.

**The number of neutrino types.** The $Z^0$ decays "democratically" into any possible channel. Thus the width of the $Z^0$ increases with the number of generations (the number of neutrino types), Fig.2 [the lifetime decreases]. The combined measurement yields N=2.9841±0.0083: the number of neutrino types with masses lower than $m_Z/2$ is three and only three. This is one of the main results of LEP and SLC.

**Determination below threshold of the top quark mass.** Virtual particles affect the masses and couplings of the EW gauge bosons. In 1993 assuming the validity of the SM it was possible to deduce the top quark mass even if the top was not directly observable because it is too heavy (it was observed at Fermilab in 1995). The precision LEP measurements together with precise theoretical calculations allowed to determine accurately the t quark mass: it was a *discovery below threshold* [today the t quark mass determined below threshold is 171±10 GeV, to be compared with 174±5 GeV measured at Fermilab].

**$m_H$.** The method used below threshold for the t quark was also used for the Higgs boson. Unfortunately the quantum corrections introduced by the $H^0$ are logarithmic, thus not as sensitive as those from the t quark. One could only establish that the Higgs boson mass had to be smaller than 200 GeV [95% C.L.], see Fig.3. It is an important constraint since in the SM the Higgs mas was only confined to be between 1 GeV and 1 TeV. The comparison with the direct searches, also shown in Fig.3, will be discussed later [5].

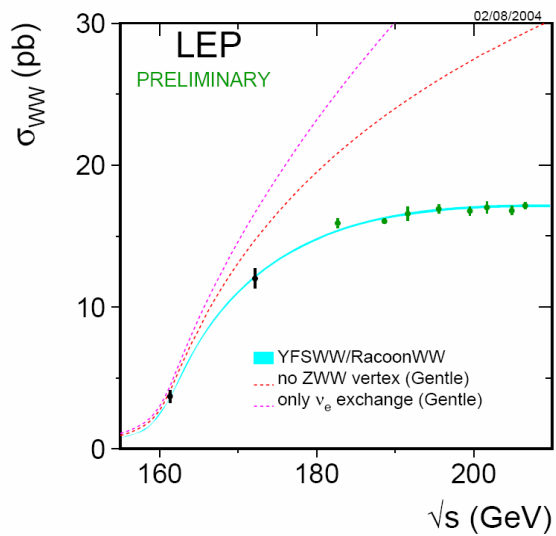

Fig.4. The total cross section for W-pair production at LEP2. The experimental data are compared with the SM prediction and also with the predictions assuming only $\nu_e$ exchange and no ZWW coupling (both excluded by the data).

**Precision measurements at LEP2.** The study of the reaction $e^+e^- \rightarrow W^+W^-$ allowed precision measurements of the W mass and the proof of the *existence of the triple bosonic vertex ZWW*, required by the SM, Fig.4 [6].

In the early LEP analyses one assumed lefthanded massless Dirac neutrinos and assumed separate conservations of the electron, muon and tau leptonic numbers. The presence of neutrino oscillations [7] forces to make changes: only the total lepton number $L=L_e+L_\mu+L_\tau$ seems now to be conserved, and one should include some right handed neutrinos. It is possible that these changes may be included in the SM, but it is also possible that the evidence for neutrino masses indicate physics beyond the SM [1].

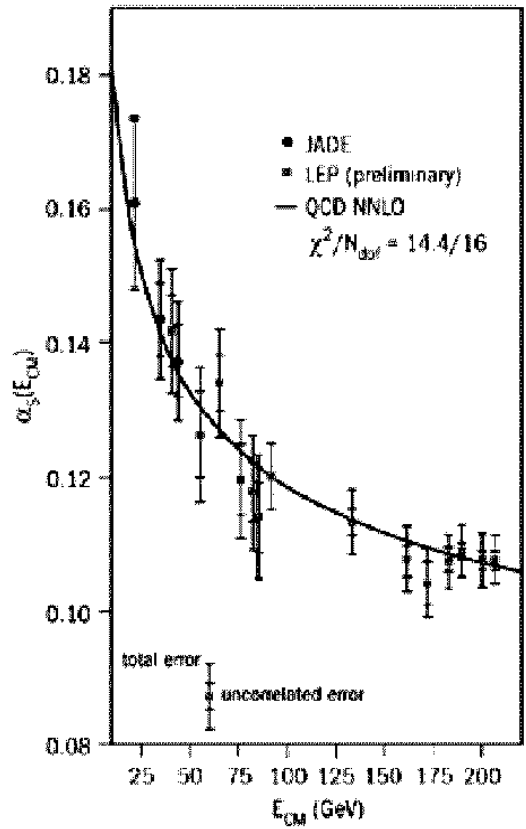

Fig.5. The running of the strong coupling constant.

## 3. QCD

Since the $Z_0$ decays predominantly into $q\bar{q}$ pairs, it yields a clean data sample with which to test quantum chromodynamics (QCD), the theory of the strong interaction. The $q\bar{q}$ pair is not observed directly, but it gives rise to two opposite jets of hadrons. Before fragmentation one of the quarks may radiate a gluon by a process similar to bremsstrahlung, yielding 3 jets of hadrons. The ratio of the number of 3-jets to the number of 2-jets is one way of measuring $\alpha_s$, the strong coupling constant.

Multihadron production in $e^+e^-$ annihilations proceeds via 4 phases [8].

1. The initial $e^+e^-$ pair annihilates into a virtual $Z^0/\gamma$, which decays into a $q\bar{q}$ pair; a $\gamma$ may be emitted by the initial $e^+$ or $e^-$. The production of the $q\bar{q}$ pair, described by the EW perturbative theory, occurs at distances of ~$10^{-17}$ cm.
2. In the second phase the q (or $\bar{q}$) radiates a gluon, which may then radiate another gluon (yielding a 3-gluon vertex), or may radiate a $q\bar{q}$ pair. This phase, described by perturbative QCD, occurs at distances of ~$10^{-15}$ cm.
3. Quarks and gluons hadronize (at distances of ~1 fm) into colourless hadrons.
4. In the 4$^{th}$ phase (described by models) the produced hadrons decay via strong or EM interactions; b-hadrons decay via WI with lifetimes of ~ $10^{-12}$ s.

**The Strong Coupling constant $\alpha_s$.** The coupling constant of the SI is a fundamental parameter which was precisely determined at LEP from many types of measurements. The experiments estabilished also the *flavour independence* of $\alpha_s$ and *the running of $\alpha_s$*, that is its decrease with increasing energy, Fig.5 (Asymptotic freedom) [9]. At the $Z_0$ mass the value of the strong coupling constant is $\alpha_s(m_Z)=0.1176\pm0.009$ : also this measurement is now a precision one!

Many other QCD studies have been made: the Colour Factors, the Physics of Heavy Flavours, the difference between hadron jets originated from quarks and from gluons [10]. Among the many phenomenological studies we may single out the study of Bose-Einstein and Fermi-Dirac correlations and the establishment of the dimensions and shapes of the hadron emission regions [11].

### 4. New particle searches

The SM has intrinsic inconsistencies and too many parameters. Many searches for new physics beyond the SM have been performed [12].

**SUSY particles.** In supersymmetric models, each particle has a SUSY partner whose spin differs by half a unit. A new multiplicative quantum number, R-parity is +1 for SM particles, -1 for SUSY partners. If R is conserved, sparticles are produced in pairs and decay to the lightest sparticle (LSP), which is the lowest mass neutralino. In the Minimal Supersymmetric Standard Model (MSSM), sparticle masses and experimental limits depend on 5 parameters.

Higgs bosons. In the MSSM one has 5 Higgs bosons, $h^0$, $H^0$, $A^0$, $H^+$, $H^-$; the neutral ones are searched for with methods as for the $H^0_{SM}$ and limits are at the same level.

Charginos. The SUSY partners of the $W^\pm$ and of the $H^\pm$ mix to form 2 mass eigenstates for each sign, the charginos. Present limits are $m_{\tilde{\chi}^\pm}$ >103.5 GeV.

Charged sleptons. Each lepton has 2 scalar partners, the right and left-handed sleptons. They could be pair produced through s-channel $\gamma/Z^0$ exchange or t-channel neutralino exchange. Mass limits: $m_{\tilde{e}}$ >99.4, $m_{\tilde{\mu}}$ >96.4, $m_{\tilde{\tau}}$ >87.1 GeV.

Scalar quarks. The decay modes $\tilde{t} \to c + \tilde{\chi}^0_1$ yield $m_{\tilde{t}}$ >95 GeV.

Neutralinos. The $\tilde{\gamma}, \tilde{Z}^0, \tilde{h}^0, \tilde{H}^0$ mix to form 4 mass eigenstates, the neutralinos. They may be pair produced through s-channel Z exchange or t-channel electron exchange. The lowest mass neutralino is unobservable at LEP; one looks for $\tilde{\chi}^0_1$ production in $e^+e^- \to \tilde{\chi}^0_1\tilde{\chi}^0_1\gamma$ or $\to \tilde{\chi}^0_2\tilde{\chi}^0_1$ with $\tilde{\chi}^0_2 \to \tilde{\chi}^0_1 l^+l^-$. The limit for the neutralino mass is ~40 GeV. Better limits exist for specific values of the SUSY parameters. The lowest mass neutralino may be a component of the *Dark Matter*.

R-Parity violation. If R is violated, sparticles may be produced singly; there are no constraints on the nature and stability of the LSP (if it has a large lifetime it crosses the whole detector). Limits are given in the context of specific models.

**Excited fermions. Compositeness.** Composite models predict the existence of excited fermions, $F^*$, with the same EW couplings to the vector bosons as the fermions. They may be produced in pairs or singly. For photonic decays the final states involve two leptons and two photons; for neutrinos, the final states involve 2 $\gamma$ plus missing energy/momentum. Present limits for singly produced excited fermions are ~102 GeV.

**Leptoquarks.** Leptoquarks (LQs) are predicted in models which explain formally the symmetry between quarks and leptons; they may be produced in pairs and each LQ decays into lepton + quark. At LEP present mass limits are >100 GeV.

**Heavy charged and neutral leptons.** Searches for long-lived charged (neutral) heavy leptons, $e^+e^- \to L^+L^-$ ($e^+e^- \to N_1\overline{N}_1 \to lW\overline{l}W$) used the central detectors and dE/dx measurements. Some searches for $L^0$'s we made assuming $e^+e^- \to L^+L^-$, $L^\pm \to L^0 W^\pm$.

**Fast heavily ionizing Dirac Magnetic Monopoles** have been searched for directly, $e^+e^- \to M\overline{M}$, using nuclear track detectors or central detectors [12].

## 5. Historical and sociological aspects

In the last 50 years there were great changes in the organization and structure of particle physics experiments. In the 1950's the standard experiment was small and was performed by a small group of physicists, students and technicians from a single University. Experimentation at higher energies forced the concentration of experiments in large national and international labs, where accelerators were available. In the 1960's the bubble chamber experiments started to be performed by collaborations of few groups from different Institutions. In the 1970's the same trend was present in counter experiments. The experiments at LEP required another step, with tens of groups and hundreds of physicists and engineers, with interconnections at the national and regional levels (future experiments at the LHC require hundreds of groups and thousands of physicists, with interconnections at the world level). In the following we discuss some aspects of the sociology of the LEP experiments and their changes with time [13].

**Preliminary workshops. The approval of the experiments.** LEP was approved in '81. Before and after approval many physics workshops were held at CERN and in different countries. After the General Meeting in Villars, Switzerland, the collaborations started to form and prepare Letters of Intents, which were presented in 1982 to the CERN Director General and to the newly formed LEP Committee (LEPC). Later followed the Proposals, which were approved by the LEPC and the CERN Research Board. Every group of each collaboration had then to obtain the financial support from the National Research Institutions. In '83 the construction of LEP and of the 4 detectors started. This first period was evaluated to be positive and stimulating by a survey ECFA (European Committee for Future Accelerators).

**Experiment construction.** Each experiment had a $4\pi$ general purpose detector, with many subdetectors and hundreds of thousand electronic channels. While many young physicists were happy to construct equipment, others feared the lack of physics papers during the long construction period.

**The first physics results.** In august '89 the first beam became available: there was a strong competition among the experiments to observe the first event. In the subsequent runs, the shape of the $Z^0$ resonance was measured and this lead to one of the most outstanding results: there are 3 and only 3 types of neutrinos. The groups were busy completing and commissioning their detectors. During this period the number of young physicists at CERN increased considerably, and everybody was very active. It was a very exciting period.

**Detector improvements.** The available forward detectors allowed luminosity measurements (using the forward $e^+e^- \to e^+e^-$ cross section) to few %. In order to fully exploit the accelerator and the detectors, precision "Luminometers" were designed and built: they allowed measurements to better than 0.1%. It was also essential to compute the radiative corrections of forward Bhabha elastic scattering to ever increasing precision: there was a healthy collaboration between theorists and experimentalists to reach the desired goal. The final measurements were made with impressive precision, much better than expected.

**Organizational structure of the collaborations.** In the large LEP collaborations there was a need of an elaborate organization, which took a long time to set up. Each collaboration had a spokesman, physics coordinators, a governing body, a sort of parliament, project leaders, subdetector committees and a financial review committee. The governing body (Executive, Steering Committee, ...) consisted of a small number of physicists, with availability of experts when needed. The Collaboration Board with the group leader of each Institution, had the last word on most items.

**From LEP1 to LEP2.** From august '89 till the middle of '95, LEP was operated at energies around the $Z^0$ peak. All fields of research benefited from the increased luminosity. From 1994 all groups started improvements in order to be ready for higher energies. Among the improvements it is worth recalling the longer and more refined vertex detectors, designed to improve the performance of Higgs searches.

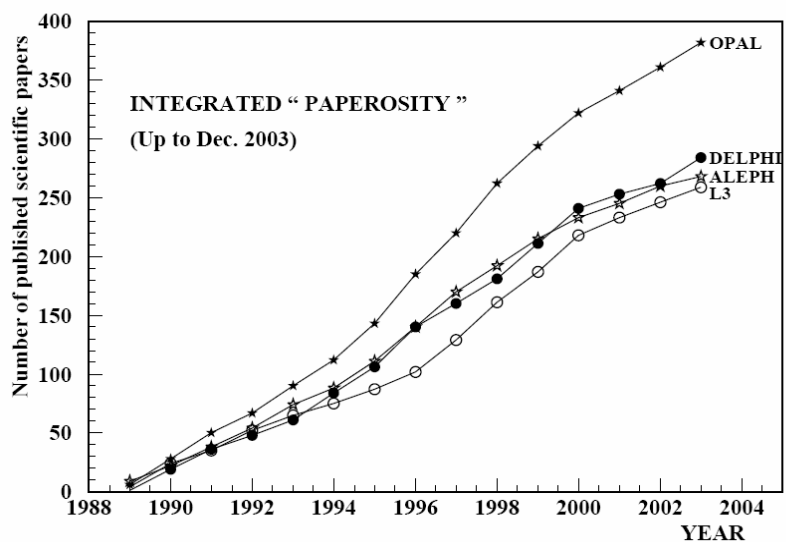

Fig.6. The integrated number of scientific papers published by each LEP collaboration from 1989 to 2003 (courtesy of Fabrizio Fabbri).

**Paperosity.** In the 11.5 years of LEP operation, 1020 scientific papers were published by the 4 collaborations, which totally included about 1730 authors. Fig.6 shows the integrated number of papers published until December 2003. Table 2 compares, for experiments at different colliders, the average number of authors and the ratio <R> = [number of papers/number of authors]. It is difficult to make a comparison since the duration of the experiments was different and the table does not include quality nor discoveries. But one can state that the LEP groups fare well in the comparison.

| Collaboration | Average number of authors | <R> |
|---|---|---|
| LEP | 330 – 550 | 0.5 – 1 |
| CDF - D0 | 400 – 500 | 0.4 - 0.7 |
| H1 – ZEUS | 350 – 450 | 0.2 – 0.4 |
| UA1 – UA2 | 65 – 150 | 0.33 |

Table 2: Various large high energy collaborations, their approximate average number of authors and the ratio <R> = [number of papers/number of authors] [11].

**Visibility.** For graduate students and young researchers it is important that their work be properly recognized. Visibility is not evident from papers with hundreds of authors. But each researcher may find his proper place inside a collaboration because of the fragmentation of responsabilities connected with the realization, operation, maintenance of complex equipments, and even more in physics analyses. Inside a collaboration there were presentations in working groups and in collaboration meetings, and also refereed and not refereed internal notes. Outside, there were presentations to conferences and invited papers. This may favour physicists who perform physics analyses, but there are also many technical workshops. From the results of the ECFA enquiry it seems that active young physicists can find a proper recognition, even inside large collaborations.

**Scientific computing.** There was on-line computing and off-line reconstruction of events, Monte Carlo studies and physics analyses. On-line computing was mostly done by clusters of Vax-stations; changes were few since it is difficult to do them in running experiments. Off-line computing had fast changes following the trend in the computing field. In 1983 the off line analyses were done by "large computers" of the type IBM 370/168 (referred to as "one unit"). At that time CERN promised to each experiment the availability at CERN of 3 computing units. At the end of LEP each experiment had a capacity of more than 1000 computing units! This change was done in steps, using new clusters of working stations. During the same time the memory space went from MegaBytes to GigaBytes, the interconnectivity improved dramatically and CERN made its most important invention: WWW, the World-Wide-Web.

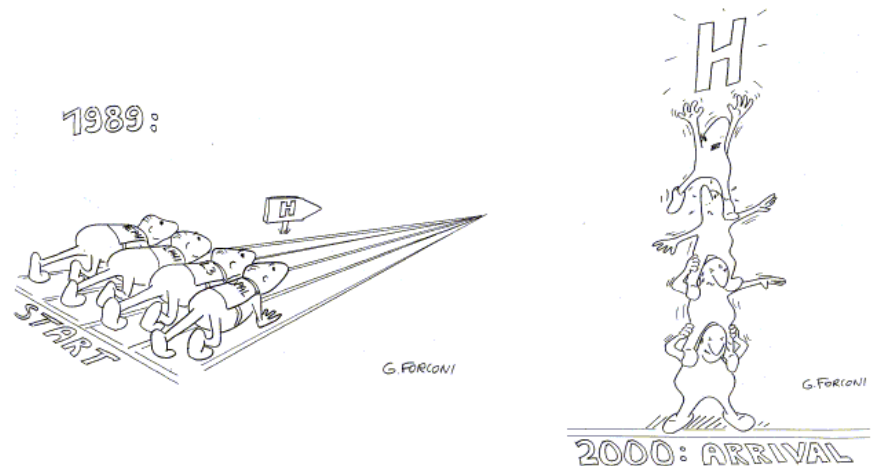
Fig.7. The Higgs search at LEP (courtesy of G. Forconi).

**LEPC.** It played an important role in all stages of the experiments. It came to an end at the end of year 2000: was it a glorious end? Did it investigate thoroughly the possibility of buying extra RF cavities, when it was still possible? An energy increase could have been important for the SM Higgs search.

Estimates of the Higgs boson mass came from precision EW data and from direct searches. While at the beginning of LEP, the 4 experiments raced one another in the search for the Higgs, at the end they combined their results as indicated in the cartoons in Fig.7. The combination of data from the 4 experiments became a standard procedure in most fields of research. It allowed to cross check data and obtain more precise results.

**The secretariats.** Each experiment had an efficient open door secretariat which provided scientific and bureaucratic information, and was called to solve every possible problem.

**Sport.** The 4 experiments participated with great enthusiasm in the sport life at CERN. Each experiment had several race teams for the annual CERN relay and road races of all categories (seniors, veterans, ladies, open, etc...). The team and sport spirit were at their best: people were happy also when they won the "random prize"!

**Love affairs.** In a large collaboration it is normal to have love affairs among collaborators, and even among members of different experiments! The sociology of "love affairs" followed the changing life pattern during the 15 years of the experiments. In the early 80's the word "fiancé" was a fine and used word, while later it almost disappeared and other terms were used, like partner, boyfriend, girlfriend, etc. There were "new experiences", course changes, encounters, new encounters and very few marriages. At the start of the new millennium there was an increase in the number of marriages, but "baby production" remains limited, much too low to compete with that in the developing countries!

## 6. Conclusions

The study of $e^+e^-$ interactions at LEP provided many important results [14,15], like the 3 neutrino generations, precision determination of the electroweak parameters and of the strong interaction parameters, the running of the strong coupling constant, precise measurements of the lifetimes of short lived particles from b and τ decays, the existence of the triple bosonic vertex, precise measurement of the W mass, the determination below threshold of the mass of the top quark, possible indications of the Higgs mass, the physics of heavy flavours [16], (b, τ), etc. It may be worthwhile to stress that the precision reached in most measurements was much better than what anticipated, and that the Particle Data Book is full of LEP results. But no new surprice was found.

One should not neglect the very large number of Diploma, Laurea and PhD theses with data from LEP, and the strong impact of LEP on the public understanding of science.

It seems that most physicists involved in one of the 4 experiments considered the LEP experience an exciting experience, in particular when they were obtaining the most interesting physics results.

**Acknowledgements.** We thank many colleagues from LEP, the LEP experiments, theoreticians, the members of the Bologna team, the technical staff and B. Poli, F. Fabbri, Y. Becherini, A. Casoni for their collaboration.